\documentclass[reprint,prb,twocolumn,showpacs,superscriptaddress,aps]{revtex4-1}

\usepackage{graphicx}
\DeclareGraphicsExtensions{.png,.jpg,.eps}

\usepackage{hyperref}
\usepackage{xcolor}
\usepackage{amsmath}

\newcommand{\redmark}[1] {{#1}}

\begin{document}

\title{\redmark{Landau Levels in 2D materials using
Wannier Hamiltonians obtained by first principles}}
%
\author{J. L. Lado}, 
\affiliation {International Iberian Nanotechnology Laboratory (INL),
Av. Mestre Jos\'e Veiga, 4715-330 Braga, Portugal
}
\author{J. Fern\'andez-Rossier} 
\affiliation {International Iberian Nanotechnology Laboratory (INL),
Av. Mestre Jos\'e Veiga, 4715-330 Braga, Portugal
}
\affiliation{
 Departamento de Fisica Aplicada, Universidad de Alicante, 03690 San Vicente del Raspeig, Spain
}



\begin{abstract}
We present a method to calculate the Landau levels and the corresponding edge
states of two dimensional (2D) crystals 
using as a starting point their electronic structure as obtained from standard density functional theory (DFT). 
The DFT Hamiltonian is represented in the basis of maximally localized Wannier
functions. This defines a tight-binding Hamiltonian for the bulk that can be
used to describe other structures, such as ribbons, provided that
atomic scale details of the edges are ignored. 
The effect of the orbital magnetic field is described using the Peierls
substitution in the hopping matrix elements.   Implementing
this approach in a ribbon geometry, we obtain both the Landau levels and the
dispersive  edge states for  a series of 2D crystals, including graphene, Boron Nitride,  MoS$_2$, Black
Phosphorous, Indium Selenide and MoO$_3$. 
Our procedure can readily be used in any other 2D crystal, and provides an
alternative to effective mass descriptions.

\end{abstract}

\maketitle

\section{Introduction}

The motion of electrons in two dimensions under the influence of a
perpendicular magnetic field $B$ results in a discrete spectrum of energy levels, associated to the
bound closed cyclotron orbits  expected within the  
  classical picture.  In the case of
Schrodinger electrons, the discrete spectrum was first calculated by
Landau,\cite{landau1962} that showed that $E_n =\hbar \omega_0
\left(n+\frac{1}{2}\right)$, where $n=0,1,$
are integer numbers and $\hbar \omega_0 =\frac{eB}{m}$,  where $e$ and $m$ are the charge and mass of the electron.
   The concept of Landau levels  (LL)  is also useful in
systems  for which the effective mass approximation is a good description of
the relevant energy bands, such as semiconductor two dimensional electron
gases.\cite{AndoRMP}      With the 
observation of the unconventional quantum Hall effect in
graphene,\cite{novoselov2005,Zhang2005} it was soon realized that the spectrum
of quantized
levels was different from the usual Landau quantization, and they would rather
scale as $E_n ={\rm sgn(n)}v_F\sqrt{2 e \hbar B |n|}$, with $n=0,\pm1, \pm2,
..$
and $v_F$ being the Fermi velocity of graphene carriers.  This unconventional
spectrum of Landau Levels 
can easily be obtained using the $kp$ 
effective mass Hamiltonian for graphene, isomorphic to  the celebrated Dirac Hamiltonian. 

\begin{figure}[t!]
 \centering
                \includegraphics[width=.5\textwidth]{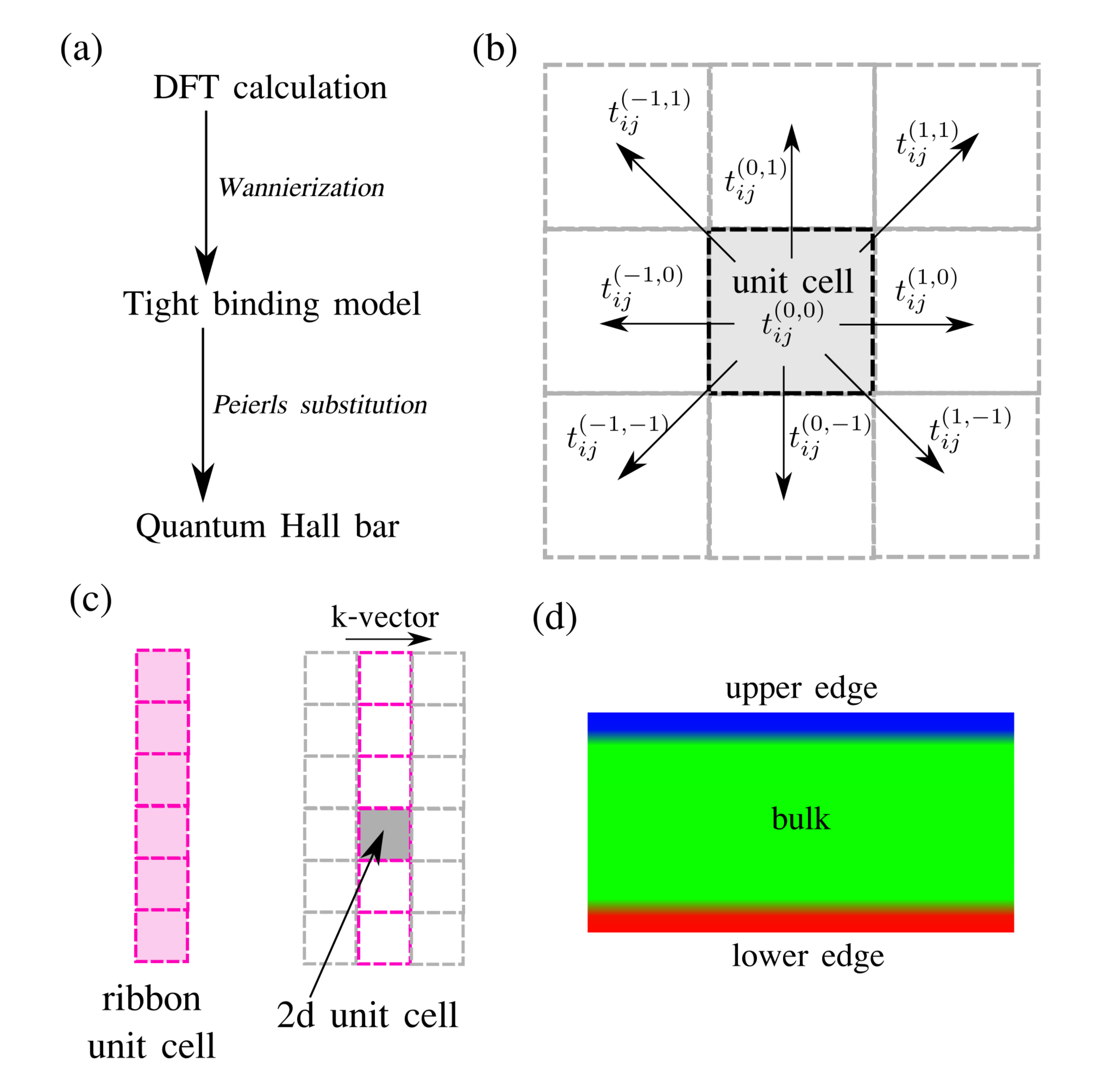}

\caption{(a) Diagram of the workflow to calculate
Landau levels for an arbitrary material using the Wannierization
procedure. (b) Sketch of the meaning in real space of the different
hopping matrices in a 2d system. From those matrices,
the Hamiltonian of the new unit cell
for a quantum Hall bar can be easily built
as shown in (c). Panel (d) shows the color code that will be used
afterwards, red and blue for the edges, and green for the bulk.
}
\label{fig:sketch}
\end{figure}

The physics of the magnetic quantum oscillations\cite{gillgren2014gate,li2015quantum,tayari2015two} and, in the extreme quantum
limit,  the quantum Hall effect of a given system, are often best described in
terms its Landau levels.  
The quest for samples with increased mobility has finally led to the observation
of the quantum Hall effect in few layer
Black Phosphorous,\cite{li2016} in thin film
transition metal disulfides
\cite{wu2015} and monolayer WSe$_2$\cite{PhysRevLett.116.086601} 
which provides an experimental motivation for this work. 
  Importantly, the properties of Landau levels can be dramatically different
depending on the symmetry of the crystal. Thus,  hexagonal crystals such as
graphene\cite{zheng2002} and MoS$_2$\cite{Niu2013}  have Landau levels that can be accounted
for in terms of two sets  of Dirac electrons,  one per valley, whereas  in the
case of black phosphorous the spectrum is similar to the more conventional case of
Schrodinger LL.\cite{pereira2015,yuan2015}  Dirac  LL  are dramatically different from Schrodinger LL,
as they come  in 3 groups, electron-like, hole-like and the intriguing $0$
Landau level.  Whereas electron and hole  Dirac-LL have a twofold valley
degeneracy,  the $0$ LL breaks valley symmetry.

The conventional procedure to calculate  the quantized levels of electrons in
two dimensional  systems, either quantum wells or 2D crystals, requires to
derive the right $kp$  effective mass Hamiltonian,\cite{saito1998physical,katsnelson2012} and solve
the Schrodinger equation, replacing
$\vec{p}$ by $\vec{p}-\vec{A}$.  The effective
mass approximation is well suited for this task because the  magnetic length
$l_B =\sqrt{\frac{\hbar}{eB}} = 25 \text{nm} /\sqrt{B\text{[Tesla]}}$ 
is much larger than the crystal  period for
laboratory scale
magnetic fields.
When applied to an infinite two dimensional system without
edges, this approach yields a set of dispersion-less LL, very often after a
simple analytic calculation. The determination of the dispersive edge states
requires to solve the equation without translational invariance in the
direction perpendicular to the edge,\cite{Halperin1982,breyfertig} which
normally requires numerical solution.    

The  implementation of the $kp$ method  can become impractical 
in some situations, such as in-plane 
heterojunctions of 2D crystals, or in
general, whenever the $kp$ Hamiltonian for bulk is not known. 
In these situations,  the calculation of both  the bulk LL and the edge 
states could be done if 
a tight-binding Hamiltonian for the system is known.  In that case, the effect
of the magnetic field is included by  doing the so called Peierls
substitution,\cite{saito1998physical}  that consists in 
replacing the hopping $t_{1,2}$ by
$t_{1,2} e^{i \Phi_{1,2}}$, where $\Phi_{1,2}= \frac{e}{\hbar}\int_1^2
\vec{A}\cdot d\vec{l}$ is the circulation of the vector potential $\vec{A}$
associated to the magnetic field.   This strategy is very often used in the
case of graphene, for which a straightforward tight-binding Hamiltonian is
available,\cite{RevModPhys.81.109}
and permits to compute the  edge states, as
well as interface  states in the case of in-plane heterojunctions.
\cite{Lado2013} However, tight-binding models are not always available.  

Here we propose a constructive  approach to obtain both  the quantized levels, edge states and interface states
 valid for a very wide variety of   2D crystals  and Van der Waals
heterostructures.
  Our strategy consists in deriving a tight-binding Hamiltonian for a given 2D
crystal starting from DFT calculations, using the well tested Wannierization
method.   
\cite{PhysRevB.56.12847,mostofi2008wannier90,PhysRevB.65.035109,PhysRevB.69.035108,RevModPhys.84.1419,PhysRevB.74.235111}
This procedure allows to obtain an exact representation of the DFT Hamiltonian in a basis set
of localized orbitals, ie, a tight-binding
representation.\cite{PhysRevB.88.245436,PhysRevB.87.235420,lado2016dirac,jung2013,PhysRevB.92.161115} 
\redmark{This procedure is carried out
in the absence of magnetic field and yields
a tight binding description that captures the topology
and orbital weight of the states in the whole Brillouin zone, as opposed to
being valid close to high symmetry points, 
giving  thereby an accurate description for complex materials.
Once this is done at $B=0$,
the addition of the effect of the magnetic field using the Peierls substitution
 is straight-forward and permits to obtain both the Landau levels and the  edge
states, provided electronic reconstructions triggered by magnetic field
are not considered.}

\section{Methods}

The Wannierization procedure 
consists in a change of basis, from the Bloch basis
$|\vec k,i\rangle$ 
which are eigenstates of the Kohn Sham
Bloch Hamiltonian
$H_{KS}({\vec k}) |\vec k,\nu\rangle = \epsilon_{\vec k, \nu} |\vec k,\nu
\rangle$, into a localized
Wannier basis $|\vec n,\nu\rangle$. Here  $\nu$ stands for band index. 
The change of basis is performed by an integration
of the Bloch waves in the whole Brillouin zone, weighted
by a gauge field $U(\vec k)_{\mu,\nu}$, so that
$|\vec n,\mu\rangle = 
\frac{1}{(2\pi)^2}
\int e^{i\vec n \cdot \vec k}U(\vec k)_{\mu,\nu} |\vec k,\nu\rangle d^2 k$.
In particular, the change of basis is characterized by the
unitary field
$U(\vec k)_{\mu,\nu}$, which 
 is chosen
so that the Wannier orbitals have the smallest spread in real space,
giving rise to the so called maximally
localized Wannier functions.\cite{RevModPhys.84.1419,PhysRevB.56.12847}
The Hamiltonian in the Wannier basis is no longer diagonal,
but due to the localized nature of Wannier states, it
only couples states whose positions are close in real space,
giving rise to a sparse Hamiltonian.
This procedure
is performed in the set of bands relevant for the
low energy properties, giving rise to a Hamiltonian with a size
much smaller than the original DFT one, but reproducing the
Hamiltonian in the energy window of the Wannierized bands.

Due to the small matrix size of the Wannier 
Hamiltonian, it is possible to precisely calculate
quantities that depend strongly on the number of k-points.
This procedure has been successfully applied (among others)
to calculate optical,\cite{assmann2015woptic}
thermoelectric,\cite{pizzi2014boltzwann}
ballistic transport\cite{PhysRevB.69.035108}
and strong correlations through
dynamical mean field theory.\cite{PhysRevB.74.125120}
Here we are interested in the calculation of Landau levels and edge states
in 2D crystals and, as we discuss now,   the  DFT calculation is done for the
unit cell of the
bulk 2D crystal.
The obtained Wannier Hamiltonian
represents a two dimensional tight binding model
and has the following form
\begin{equation}
H_{2D} = \sum_{n,m}\sum_{i,j}t^{\vec n - \vec m}_{ij}
c^\dagger_{\vec m,j} c_{\vec n,i}
\end{equation}
which in reciprocal space reads
\begin{equation}
H_{2D}(\vec k) = \sum_{n,m}\sum_{i,j}
e^{i\vec k \cdot(\vec m - \vec n)}
t^{\vec n - \vec m}_{ij} c^\dagger_{\vec k,j} c_{\vec k,i}
\end{equation}
where $i,j$ are the indexes of the different orbitals in the unit cell
and $\vec n,\vec m$ 
are the vectors that label the different unit cells.
The matrices $t^{\vec n - \vec m}_{ij}$
are the hoppings between the Wannier orbitals
between the different unit cells,
\begin{equation}
t^{\vec n - \vec m}_{ij} = \langle \vec n,i | H_{KS} |\vec m,j \rangle
\label{hoppings}
\end{equation}
with $|\vec n,i \rangle$ the $i$-th Wannier orbital
in the cell $\vec n$.
In Fig. \ref{fig:sketch}b
it is shown a sketch of the meaning of those matrices in real space,
for the case of a square lattice and hopping to first neighboring cells.

Once we have the hopping integrals Eq. (\ref{hoppings}),  obtained for the
minimal unit cell 
that describes the 2D crystal, we can build a  model for a one  dimensional
slab, as schematically shown in 
Fig. \ref{fig:sketch}c.  
In the same step it is  also possible to 
add the effect of the magnetic field acting on the orbital degrees of freedom
 using the Peierls substitution. 
Assuming that the bar is infinite 
in the $x$ direction, we use
the Landau gauge $\vec A = (By,0,0)$ that 
maintains the translation symmetry of the Hamiltonian,
and modifies the hoppings $\hat t_{ij}$
of the 1d unit cell according to
\begin{equation}
\hat t^{N - M}_{ij} \rightarrow
\hat t^{N - M}_{ij} e^{i\phi^{N,M}_{ij}}
\end{equation}
where
\begin{equation}
\phi^{N,M}_{ij} = \frac{eB}{2\hbar}(x^N_i - x^M_j)(y^N_i+y^M_j)
\end{equation}
 is the Peierls phase, $B$ the magnetic field
and $x^N_{i},y^{N}_{i},x^M_{j},y^{M}_{j}$ 
the center of the different Wannier orbitals
in the cells $N$ and $M$
of the 1d system. Finally, by calculating the expectation
value of the position along the width of the ribbon for each state,
we obtain whether a certain state is a bulk LL or an edge state
(see Fig. \ref{fig:sketch}d).

Since the method uses the same matrix elements for edge and bulk atoms,  
it completely misses  the  atomic scale
reconstructions at the edges, such as dangling bonds and any other edge specific
atomic scale process. The method assumes that the edges are
identical to bulk, except for the reduced coordination. Whereas this assumption
is certainly not realistic to describe atomic scale edge properties,   it
provides a quite reliable description of both bulk LL and even the   edge
states associated to propagating modes along the boundaries of the sample,
whose localization length along the transverse direction is
given by  $\ell_B$,
much larger than the lattice constant.

\section{Results}

\subsection{Graphene}

\begin{figure}[t!]
 \centering
                \includegraphics[width=.5\textwidth]{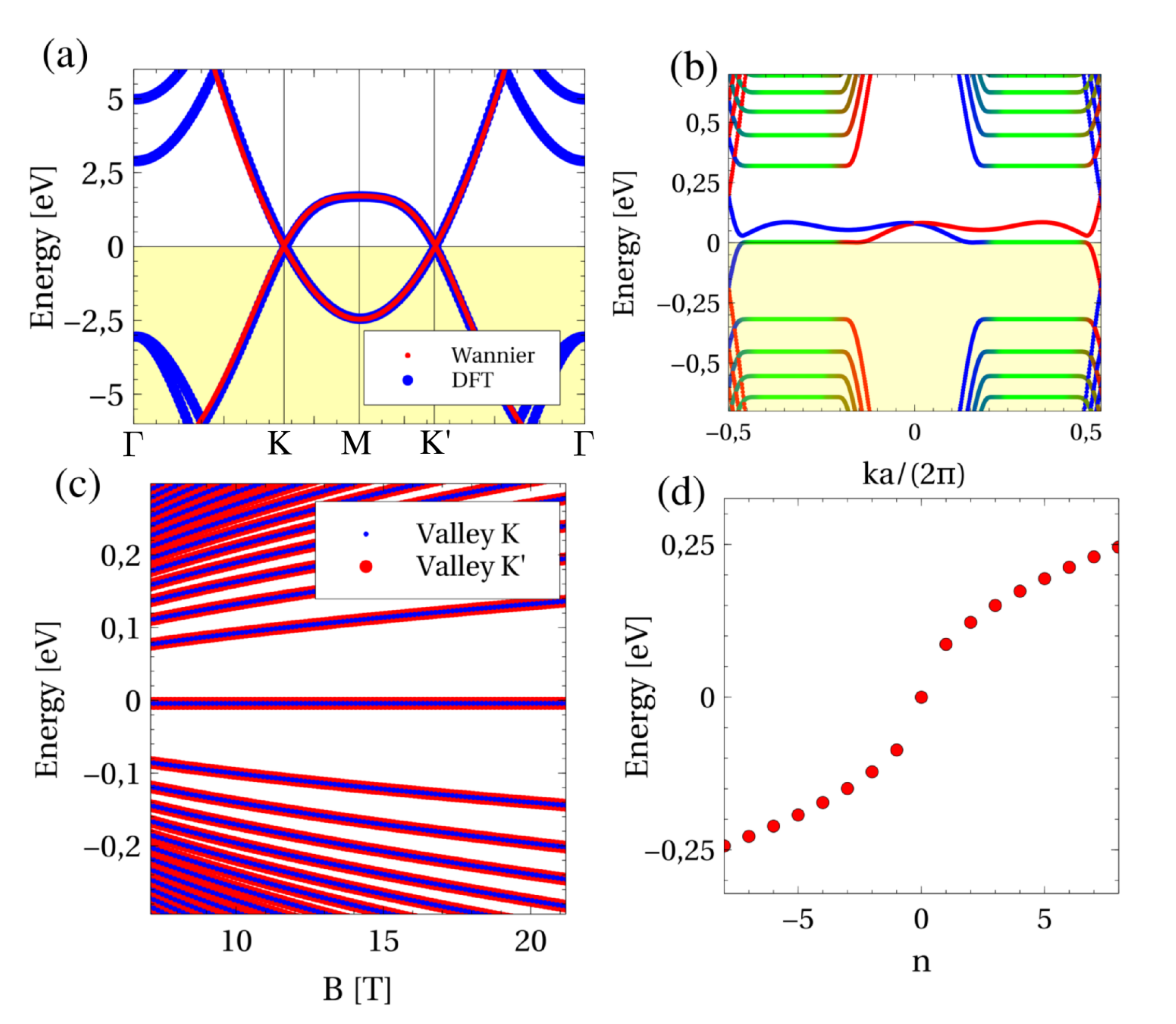}

\caption{ (a) Comparison of the band structure of graphene
as obtained by DFT and the Wannier
tight binding Hamiltonian. (b) Band structure of graphene
nanoribbon of thickness 25 nm
in the quantum Hall regime ($B$ = 120 T).
(c) Evolution of the Landau levels with magnetic field
as obtained with the Wannier Hamiltonian, and evolution
of the Landau level energy with the index for $B = 8$ T.
Ribbon thickness in (c,d) is 170 nm.
}
\label{fig:graphene}
\end{figure}

We now test   the procedure  with  graphene.\cite{RevModPhys.81.109}
In a single graphene layer, the low energy properties are
dominated by two $p_z$-like orbitals, one in each carbon atom. At higher
energies, the decoupled $p_z$ bands coexist with bonding/antibonding
$sp^2$ states. To perform the Wannierization, a frozen window of $[-1,1]$ eV
is chosen so it contains the low energy region, whereas the outer
window goes up to $[-9,+9]$ eV to capture
the whole pz manifold. The comparison between the full
DFT band structure and the Wannier band structure is shown in
Fig. \ref{fig:graphene}a.  It is apparent that the Wannierization
captures both the low energy Dirac dispersion as well
as the electron hole asymmetry which arises due to second neighbor
hopping.\cite{PhysRevB.88.165427}

Upon application of a magnetic field in a ribbon build with the previous
Hamiltonian, the familiar set of Dirac Landau levels appear. In
particular (Fig. \ref{fig:graphene}b), a
single zero Landau level per valley
shows up, that
connects with  zigzag edge states which show some dispersion due to the finite 
second  neighbor hopping.\cite{PhysRevB.88.165427}  Except for this feature,
associated to states atomically localized at the edges,  the method yields
results identical to those obtained with the standard tight-binding
model.\cite{katsnelson2012}
The color
of the bands in 
Fig. \ref{fig:graphene}b
 shows the spatial location of the state: green stands for {\em
bulk}\footnote{As opposed to being localized in the edge},   whereas red and
blue stand for top and bottom edge respectively.
We can repeat the calculation for several values of $B$ and study the evolution
of the Landau level spectra  as a function of $B$.  We  obtain  the expected\cite{zheng2002,katsnelson2012} square root behavior 
of the energy with $B$. Independently
on the magnetic field the first Landau level is always pinned
at zero energy and two fold degenerate, originating
one from each valley.
Finally,
the evolution of the LL energy for a fixed magnetic field as a function
of the LL index can be obtained in the same way, showing the expected
square root behavior $E_n \propto \sqrt{n}$. This further confirms the reliability of our method.

\subsection{Boron nitride}

Hexagonal boron nitride
\cite{gorbachev2011hunting,alem2009atomically}  (BN) 
is a two dimensional material
which is mostly known for its insulating behavior and its
extraordinary properties for acting as a 
high quality substrate
for other 2D materials.
\cite{young2014tunable,PhysRevB.93.115441,doi:10.1021/acsnano.5b01341,britnell2012atomically}
Very much like graphene,
BN  consist on a honeycomb lattice, but with
two inequivalent atoms, boron and nitrogen, in the unit cell.
Its electronic structure
is usually understood as a gapped Dirac equation
in the p$_z$ manifold, having a direct band gap,
although  recent findings suggest that
its bulk form
shows an indirect gap.\cite{cassabois2015hexagonal}

\begin{figure}[t!]
 \centering
                \includegraphics[width=.5\textwidth]{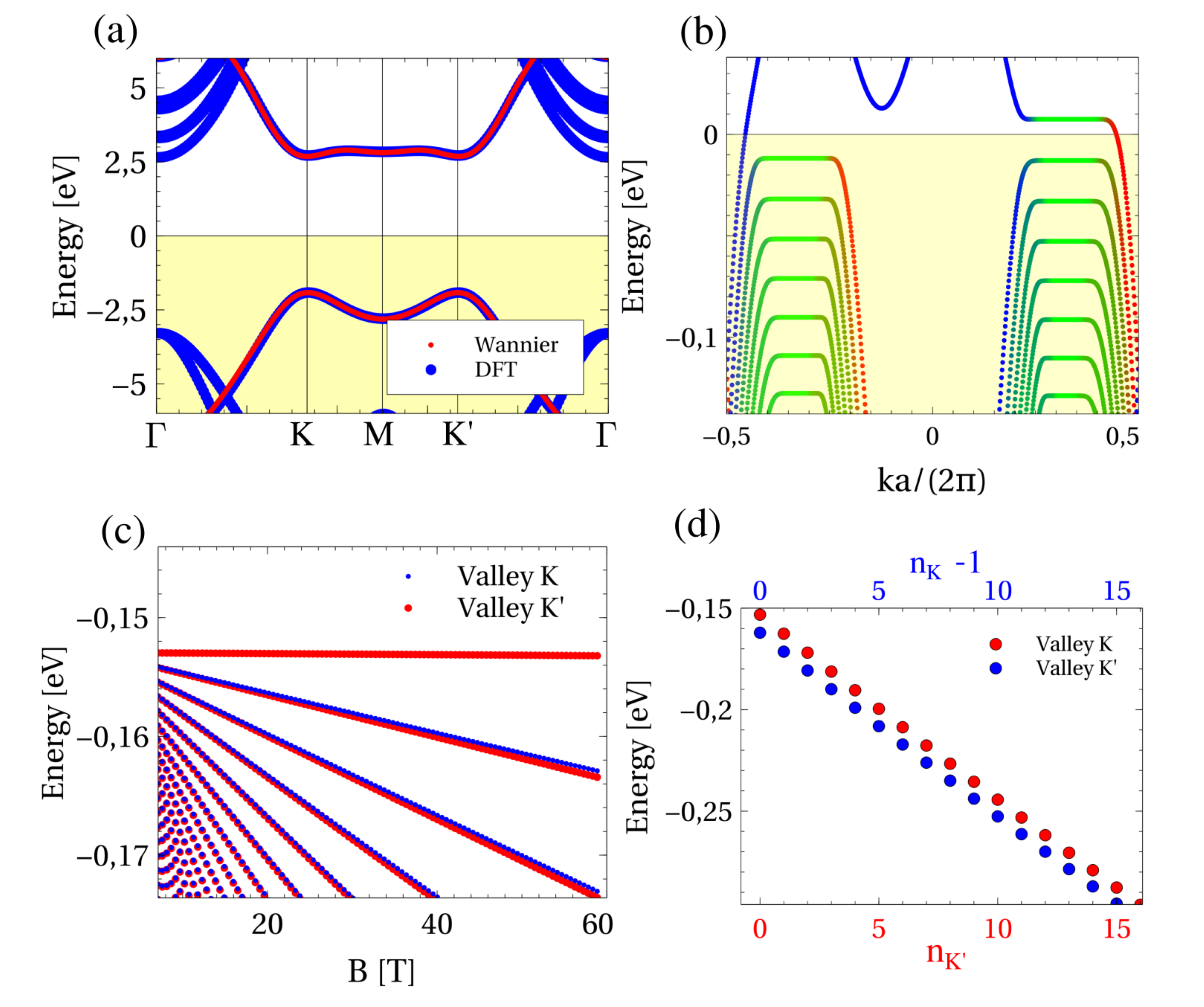}

\caption{ (a) Comparison of the band structure of monolayer
boron nitride
as obtained by DFT and the Wannier
tight binding Hamiltonian. (b) Band structure of hole doped BN
nanoribbon of thickness
35 nm in the quantum Hall regime ($B = 116$ T), focusing on the valence band.
(c) Evolution of the Landau levels with magnetic field
as obtained with the Wannier Hamiltonian, and evolution
of the Landau level energy with the index (d) for $B = 50$ T.
Ribbon thickness in (c,d) is 175 nm.
}
\label{fig:BN}
\end{figure}

The electronic structure obtained with DFT is shown in Fig. \ref{fig:BN}a,
as well as the comparison with the  bands obtained via Wannierization with
p$_z$-like orbitals of boron and nitrogen.
The low energy properties
show a strong electron-hole asymmetry, giving rise in the conduction
to a quite flat band between $K$ and $K'$,
so that in the following we will focus
on the more conventional valence band.  In figure 
Fig. \ref{fig:BN}b we show 
the Landau levels obtained with the Wannier Hamiltonian, focusing on the
valence band.
It is apparent that the spectrum is different for $K$ and $K'$ valleys
(positive and negative values of $k$ in the figure), as expected for the case
of a  massive Dirac equation.\cite{koshino2010}   In particular, the $n=0$
Landau level for holes  is only present in one of the valleys.  The scaling of
the Landau levels with
the magnetic field is shown in Fig. \ref{fig:BN}c, where it is observed
the flatness with $B$ of the 0LL, expected in a massive Dirac equation\cite{koshino2010} and the almost linear 
dispersion of the Landau Levels with $B$, expected for Dirac electrons with a large mass.

\subsection{MoS$_2$}
Transition metal dichalcogenides
are another set of 
materials that can be exfoliated
into 2D flakes and are attracting enormous interest.\cite{wang2012}
They also have a hexagonal structure, but their electronic structure is more complicated
than the one of graphene and BN, involving several $d$ orbitals of the transition metal,  and also some
contribution coming from the $p$ orbitals of the group VI
atom.\cite{PhysRevB.88.245436}
Therefore,
obtaining a Slater Koster
model or a tight binding model
by fitting to the band structure
is a very challenging task.
\cite{zahid2013generic,cappelluti2013tight,ridolfi2015tight,
PhysRevB.88.085433,kormanyos2015k,liu2015electronic}
Even if a good fitting is obtained, the fact that the
parametrized Hamiltonian reproduces the topology
and orbital character
of the original Hamiltonian has to be carefully checked.
In contrast, the Wannierization procedure
allows to get the tight binding Hamiltonian
in a single shot, reproducing both the topology
and orbital weights of the bands.

Unconventional Hall effect
in dichalcogenides
\cite{Niu2013,PhysRevB.93.035406,PhysRevB.88.125438} 
is expected due to the
Dirac-like nature of its band structure.\cite{xiao2012}
In the following, we will focus
on the case of MoS$_2$, although a similar analysis
can be applied to other 
transition metal dichalcogenides (TMD).
We study first the conduction band, for which the effect of spin orbit coupling
is much smaller\cite{PhysRevB.88.245436}   than in the valence band,
so that it can be initially neglected. In particular, the spin-orbit splitting 
in the conduction band
is  much smaller than   the Landau level splitting  for moderate values of $B$.
For the particular case of MoS$_2$,
the orbitals chosen as initial guess are the p orbitals
in S and the d-orbitals in Mo, giving rise to a 11 band Hamiltonian.\cite{PhysRevB.88.245436}

\begin{figure}[t!]
 \centering
                \includegraphics[width=.5\textwidth]{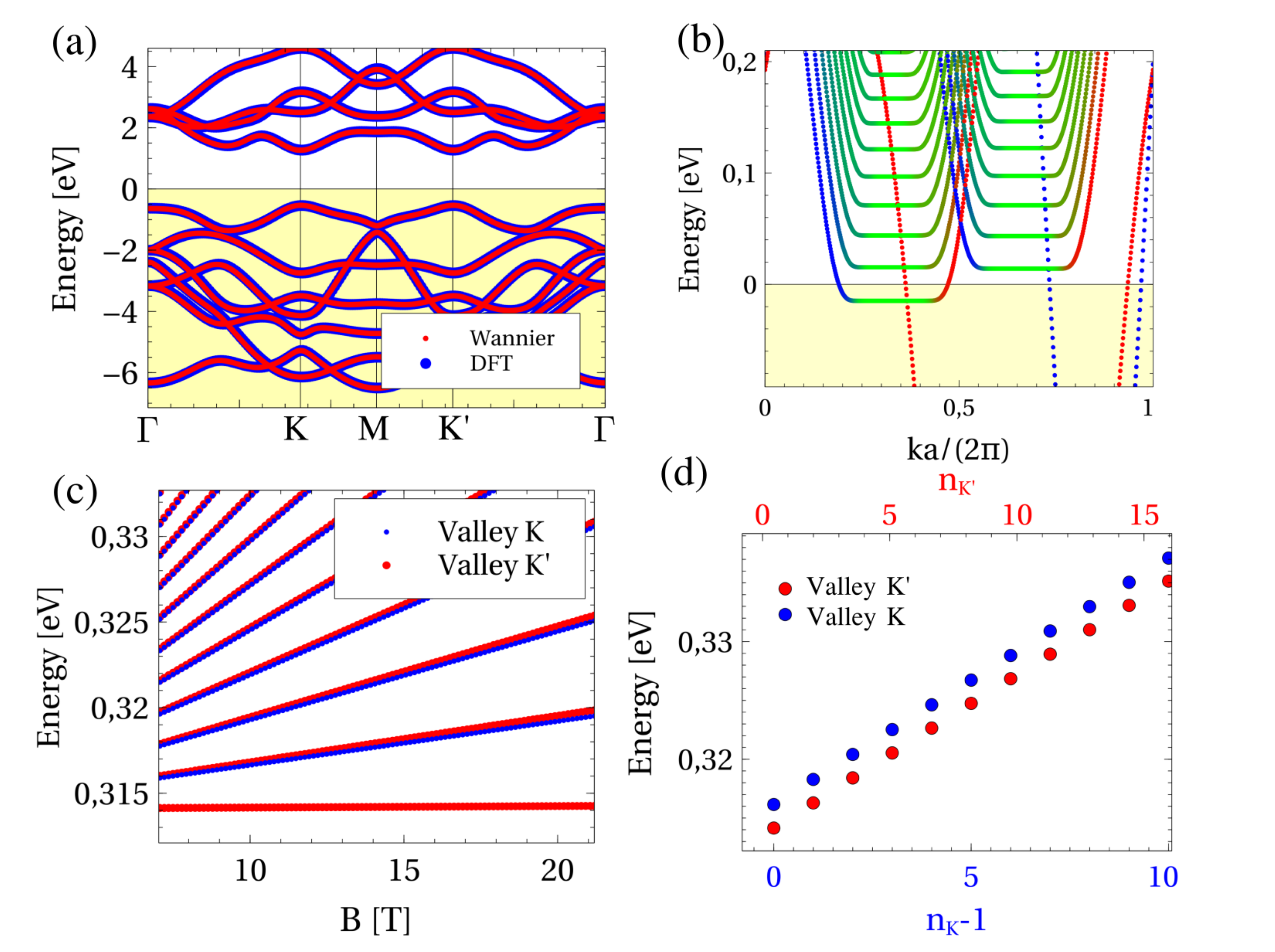}

\caption{ (a) Comparison of the
non relativistic band structure of MoS$_2$ as obtained by DFT
(blue dots) and obtained by the tight binding model derived by Wannierization
(red dots). 
(b) Band structure of a quantum Hall
MoS$_2$ bar (ribbon thickness 33 nm and  $B = 120$ T),
focusing on the conduction band,
with the color representing the spatial
position of the eigenvalue.
It is observed that
only one valley yields a 0LL.
(c) Evolution of the Landau levels in the conduction band
as a function of the magnetic field, resolved for each valley. 
(d) Dependence of the the Landau level energy with the
Landau level index n for
$B = 8 $ T as obtained from the tight binding calculation.
Fermi level in (b) is shifted with respect to (a).
Ribbon thickness in (c,d) is 220 nm.
}
\label{fig:MoS2}
\end{figure}

The comparison between the DFT and Wannier Hamiltonians obtained is shown
in Fig. \ref{fig:MoS2}a.
We emphasize that since the Wannierization
is simply a change of basis, the orbital information is perfectly
conserved between the DFT Kohn Sham states and the tight binding Hamiltonian.
With the previous Hamiltonian a quantum Hall slab can be built, that permits to compute 
the
Landau level spectra shown in Fig. \ref{fig:MoS2}b. 
Because of the lack of inversion symmetry,   the $n=0$ 
 Landau level in the conduction band is only present in one valley
as observed in Fig. \ref{fig:MoS2}b.
 \cite{PhysRevB.90.045427,PhysRevLett.114.037401}
In addition, for
Landau indexes $n>0$, 
a sizable valley splitting has been predicted,
\cite{PhysRevB.90.045427,PhysRevB.91.075433,PhysRevB.88.125438}
based on a three band tight binding model,
feature that would not be observed in the Landau levels of a
massive Dirac equation.
Including hopping up to third neighboring
cells, we find that the valley
splitting in the conduction band is rather small.
In comparison, if we only retain hopping to the first neighboring
cell, we recover the sizable intervalley splitting
predicted.\cite{PhysRevB.90.045427,PhysRevB.91.075433}
We thus conclude that the $n>0$
valley splitting in the conduction band
depends strongly on the details of the tight binding Hamiltonian
used.

\begin{figure}[t!]
 \centering
                \includegraphics[width=.5\textwidth]{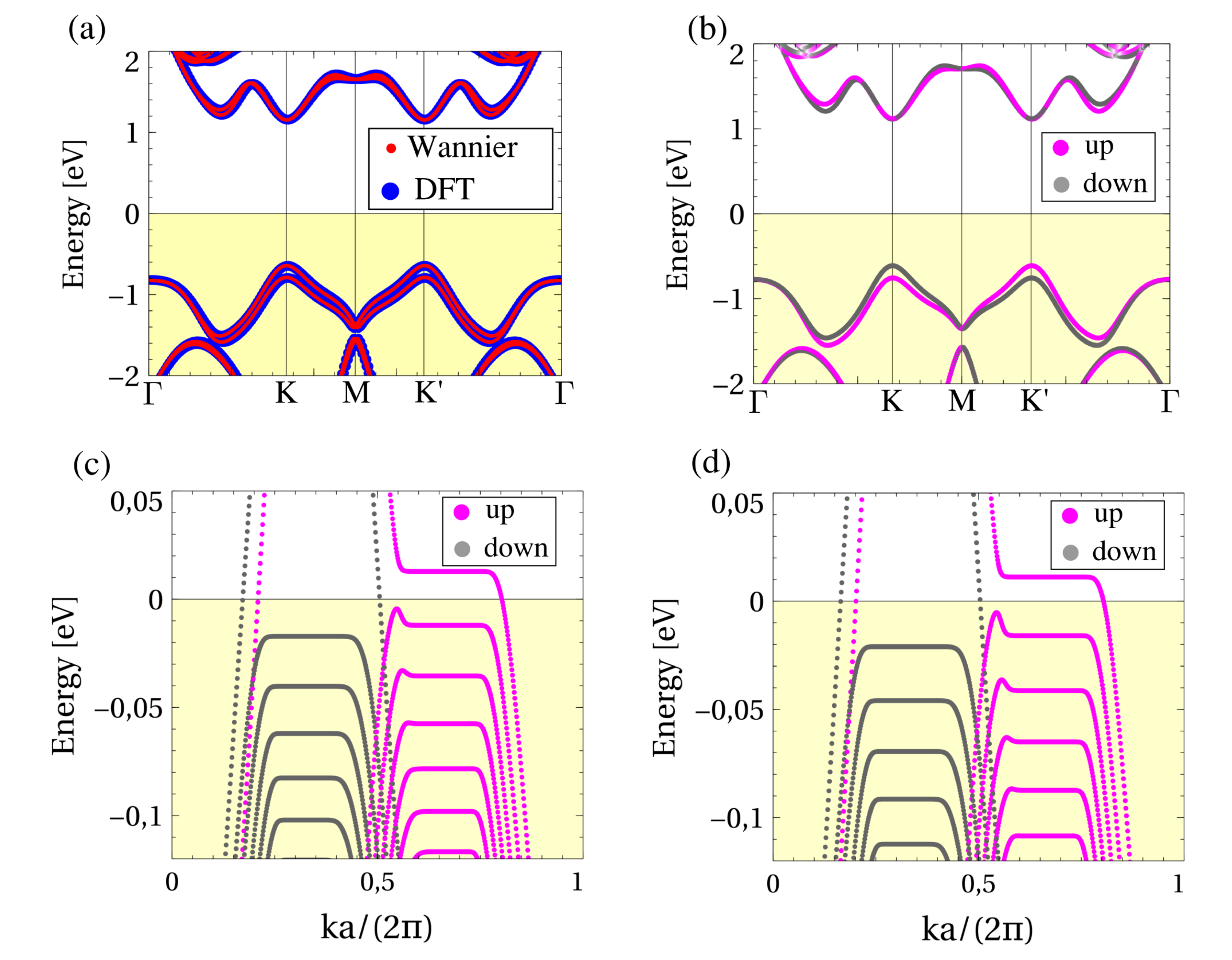}

\caption{ (a) Comparison of the
band structure of MoS$_2$ as obtained by 
a relativistic DFT
(blue dots) and obtained by the tight binding model derived by 
a relativistic Wannierization
(red dots). 
(b) Band structure obtained by adding atomic SOC to the
non relativistic tight binding Hamiltonian obtained with a spinless
Wannierization (Fig. \ref{fig:MoS2}). Panels (c,d) show
the band structure of a quantum Hall slab, focusing in the valence band,
using the relativistic Wannierization (c) and
the spinless Wannierization plus atomic SOC (d).
Colors in (b,c,d) indicate the spin flavor of the eigenvalue.
Ribbon thickness in (c,d) is 33 nm and magnetic field $B = 120$ T.
}
\label{fig:MoS2_soc}
\end{figure}

So far we have ignored the effect of 
spin orbit coupling.
We can implement the method 
used so far starting from a DFT calculation
that includes  spin orbit coupling, and  performing a Wannierization over a fully relativistic calculation.
The results of this procedure are shown  in
Fig. \ref{fig:MoS2_soc}a, and the Landau levels for the
valence band in a 
quantum Hall slab are shown in
Fig. \ref{fig:MoS2_soc}c.

Nevertheless,
 the effect
of SOC can also be captured 
without a fully relativistic Wannierization, but just
by adding an atomic-like
SOC term to the spinless Wannierization performed previously.\cite{PhysRevB.88.245436}
This will be valid as long as
the Wannier orbitals are atomic-like close to the atom\cite{PhysRevB.87.235109,PhysRevB.88.245436}
(which is where the SOC has its strongest contribution),
and provided the SOC splitting comes from the manifold where
the Wannierization was performed.
For example, in the case of graphene, the first principles SOC gap 
of 40 $\mu$eV will be reproduced
with the previous procedure
only if the
Wannierization includes d-orbitals of carbon,\cite{PhysRevB.80.235431,PhysRevB.82.245412}
whereas only inclusion of sp orbitals would yield
a gap 1 $\mu$eV for realistic SOC coupling,
\cite{PhysRevB.74.165310,PhysRevB.75.041401}
much below the actual DFT value.

The inclusion
of SOC after the Wannierization gives rise
to the following Hamiltonian
\begin{equation}
H = H_\text{spinless} + H_{SOC}
\end{equation}
where
\begin{equation}
H_{SOC} = \sum_{a\in \text{atoms}}\lambda_a 
\sum_{(i,j) \in a}(\vec L \cdot \vec S) 
c^\dagger_{i,s}c_{j,s'}
\label{wsoc}
\end{equation}
is the atomic spin-orbit coupling.
$H_{\text{spinless}}$ is the Wannier Hamiltonian
obtained from the non relativistic DFT calculation,
for the bulk or the ribbon depending on the case.

This procedure avoids having to perform
a relativistic calculation, and can be useful to approximately capture SOC
effects if a particular DFT code lacks of relativistic implementation,
if the fully relativistic calculation is computationally too expensive,
or simply to study how the band structure evolves as the SOC
is turned on.\cite{PhysRevB.88.245436} 
It is important to note that effects produced by variations of the
DFT charge density by SOC effects will not be captured, so that
small differences in dispersions and effective masses are expected.

The values of $\lambda_{Mo}$ and $\lambda_S$ 
in Eq. \ref{wsoc} correspond
approximately to the
atomic SOC in Mo and S, which was shown\cite{PhysRevB.88.245436} to be on the
order of 80 meV for Mo and 50 meV for S. The
bulk band structure obtained with the non relativistic Wannierization
plus atomic SOC following Eq.\ref{wsoc} is shown in Fig. \ref{fig:MoS2_soc}b.
It is observed that this method gives a band structure in good agreement with
the fully
relativistic one (Fig. \ref{fig:MoS2_soc}a),
in particular reproducing the valley spin splitting
in the $K$ and $K'$ points. The quantum Hall effect in this particular
system is shown in Fig. \ref{fig:MoS2_soc}c for the fully
relativistic Wannierization and in Fig. \ref{fig:MoS2_soc}d
for the non-relativistic Wannierization plus atomic
SOC. In both cases it is observed that due to the large spin
splitting in the valence valleys, the Landau levels
are spin polarized for each valley, and with a valley-dependent  spectrum for the values of the
 Landau index larger than $0$.   More importantly, only one of the valleys
shows a single Landau level, which turns the system into a quantum
spin ferromagnet upon hole doping. Although methods
\ref{fig:MoS2_soc}c,d give qualitative similar results, a small difference in
effective mass between both calculations create a small misalignment
between the levels.

Finally, it is worth to note that the previous calculations
do not include nor the Zeeman term $H_Z = 2 \mu_B \vec B \cdot \vec S$
neither the coupling to the
atomic orbital angular 
momentum  $H_A = \mu_B \vec B \cdot \vec L$,
where $\vec L$ is the atomic angular momentum,
$\vec S $ the spin, and $\mu_B$
is the Bohr magneton.
This orbital coupling is internal and it is
different from the orbital magnetization.\cite{PhysRevLett.95.137205,
resta2010electrical,PhysRevB.74.024408}
In the case of MoS$_2$, the last angular term would create
a valley dependent splitting in the valence band
that increases linearly with $B$, since
the states
in the valence band valleys are dominated by $L_z = \pm 2$ from Mo.
In comparison, its contribution in the conduction band
will be much smaller, due to the $L_z=0$ dominant character. 
The Zeeman term would create a spin splitting in both
valleys, whose effect would be more important
in the conduction band, where at large enough fields would
be able to compete with the small SOC splitting of
the conduction band.

\subsection{Black phosphorus, InSe and MoO$_3$}

In the following
we will apply the method presented to other
semiconducting two dimensional materials. In particular
we will study cases whose low energy properties
are believed to be dominated by Schrodinger-like
dispersion relations, rather than Dirac like.

We first
turn our attention to monolayer black phosphorus,
\cite{castellanos2014isolation}  a two dimensional
semiconductor that shows highly anisotropic electronic
properties due to its distorted lattice.
Its a direct gap semiconductor, with  top
of the valence  and bottom of the conduction bands
located
at the $\Gamma$ point in the Brillouin zone.

\begin{figure}[t!]
 \centering
                \includegraphics[width=.5\textwidth]{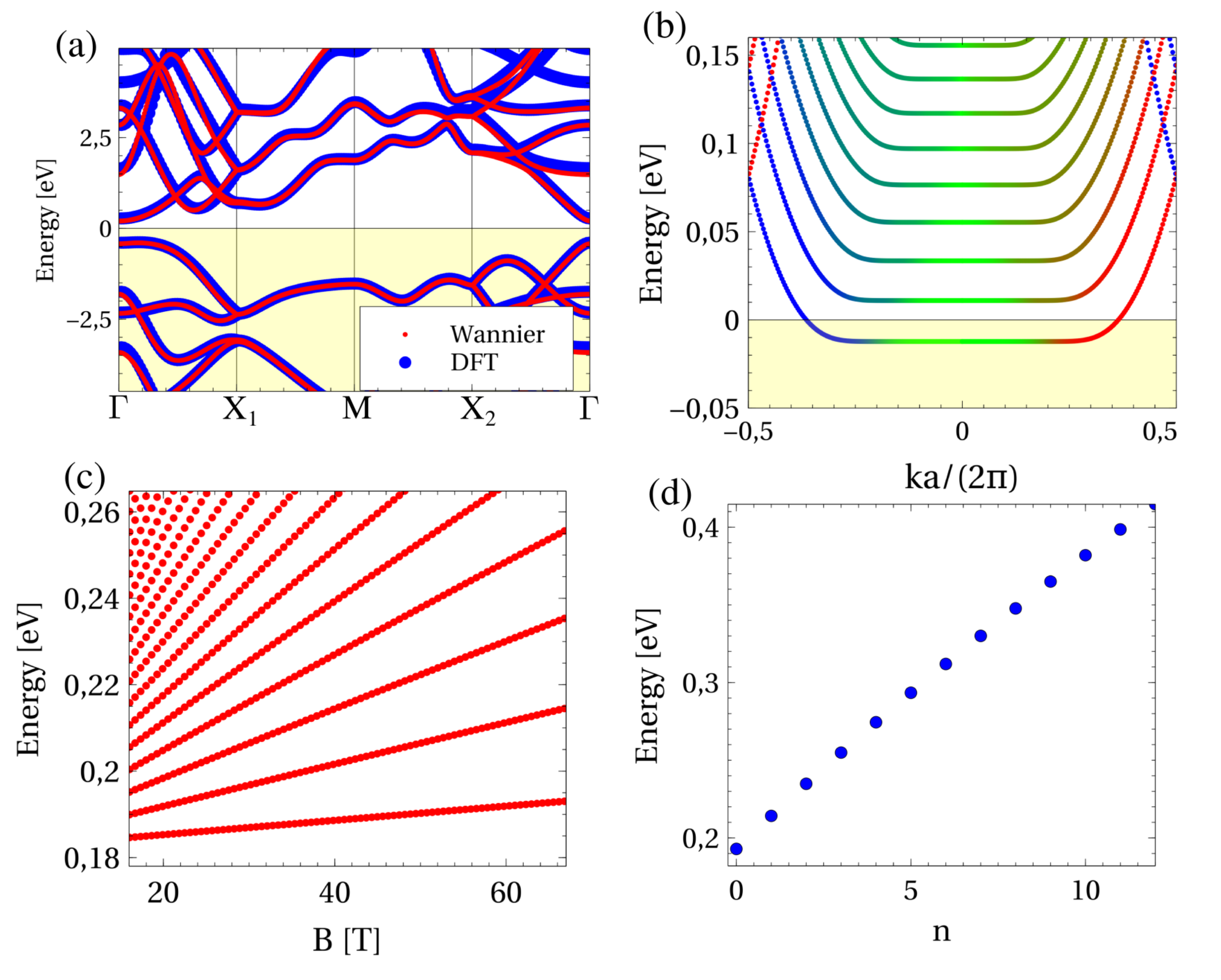}

\caption{ (a) Comparison of the band structure of monolayer
black phosphorus as obtained by DFT and the Wannier
tight binding Hamiltonian. (b) Band structure of black phosphorus
nanoribbon of thickness 90 nm
in the quantum Hall regime ($B = 70$ T), focusing on the conduction band.
(c) Evolution of the Landau levels as a function of the magnetic field.
(d) Dependence of the Landau level energy as a function of the
Landau level index. Ribbon thickness in (c,d) is 360 nm.
}
\label{fig:P}
\end{figure}

 The strong mixing between s and p orbitals
in black phosphorus, together with the
low symmetry of the unit cells turns the fitting of its electronic structure
a very challenging task if few orbitals are
considered.\cite{takao1981electronic,
rudenko2014quasiparticle,rudenko2015toward}
In order to properly capture
the electronic structure of black phosphorus,
we carry out the Wannierization with
16 Wannier orbitals , corresponding to the $s$ and $p$ orbitals of the 4 atoms of the unit cell.
The inclusion of all those orbitals gives
rise to a tight binding model that agrees well with the DFT
eigenvalues as shown in Fig. \ref{fig:P}a.
The Wannierization with the 16 orbitals
is not computationally expensive, and importantly
its construction in this way
warrants that all the orbital weights are properly captured.

Implementing this  Hamiltonian in a one dimensional 
black phosphorus slab, and using the
Peierls substitution,
we obtain the spectrum of
Landau levels\cite{zhou2015landau} 
and edge bands shown in Fig. \ref{fig:P}b.
The anisotropic nature of the low energy properties
averages out,\cite{pereira2015,zhou2015landau}
and the resulting scaling with the Landau level
index (Fig. \ref{fig:P}d) and magnetic field (Fig. \ref{fig:P}c) are
in line with those obtained using the effective
mass approach.\cite{pereira2015}


We now  implement our method for a monolayer of
indium selenide,\cite{debbichi2015two} another semiconductor
that can synthesized in 2D form.
It shows a hexagonal lattice, very much
like TMD, but with four atoms per unit cell
instead of three,
two In and two Se.
\cite{debbichi2015two,lauth2016solution,lei2014evolution}
It has an indirect gap between the conduction band at $\Gamma$
and a Mexican hat around $\Gamma$ in valence,\cite{debbichi2015two}
with a direct gap of similar value.
The gap is in the order of 1.6 eV, 
and tunable by a perpendicular electric
field.\cite{debbichi2015two}
In addition,
electroluminescence\cite{electroInSe}
as well as by confinement effects\cite{brotons2016nanotexturing}
have been recently experimentally observed.

\begin{figure}[t!]
 \centering
                \includegraphics[width=.5\textwidth]{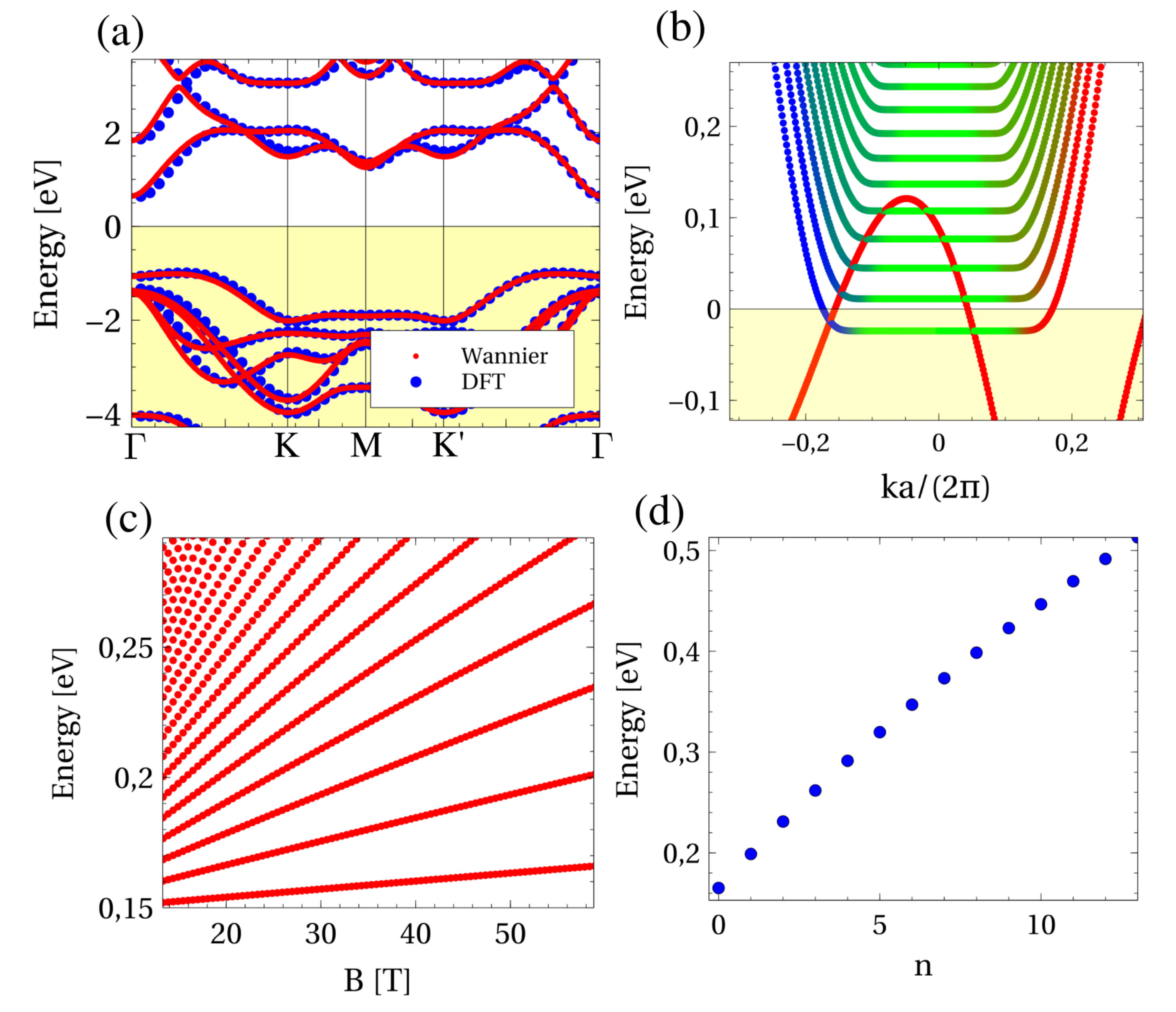}

\caption{ (a) Comparison of the band structure of monolayer
InSe as obtained by DFT and the Wannier
tight binding Hamiltonian, showing
a minimum on the conduction band at $\Gamma$. (b) Band structure of InSe
nanoribbon of 70 nm thickness
in the quantum Hall regime ($B=60$ T),
focusing on the conduction band,
showing a conventional Landau level spectra. (c) Scaling of the
Landau levels energy with the magnetic field, and with
the Landau index (d) at a fixed magnetic field.
Ribbon thickness in (c,d) is 310 nm
}
\label{fig:InSe}
\end{figure}

For the sake of simplicity,
in the following we focus on the conduction band
of InSe, that shows a minimum at $\Gamma$. 
The lack of inversion
symmetry will create a spin splitting in the band structure
once SOC is considered,
that vanishes at the  $\Gamma$ point,  a 
 time reversal invariant momenta. With that in mind, we
perform a non relativistic calculation, together with the Wannierization
(Fig. \ref{fig:InSe}a) of the InSe monolayer. The Wannierization is performed
using 14 orbitals, $s$ and $p$ of In
and $p$ of Se (each unit cell has 2 Se and 2 In).
With the Wannier Hamiltonian, the Hall bar is created (Fig. \ref{fig:InSe}b),
giving rise to a conventional Landau level spectra. The scaling
of the Landau level energy with the magnetic field
(\ref{fig:InSe}c) and with the Landau index (\ref{fig:InSe}d)
is the one of the conventional 2d electron gas. Therefore,
the conduction band of InSe is one of the cleanest examples of
Schrodinger dispersion in a 2d material, lacking of
SOC splitting or Berry curvature effects.  
In striking comparison,
the valence band will show
both SOC effects and Berry
curvature effects, since the top of the valence band is
not at $\Gamma$.


The last two dimensional material we consider is 
MoO$_3$,\cite{kalantar2010synthesis}  another semiconducting material
that can be
brought to 2D form\cite{molina2015centimeter}
giving rise to a bilayer system formed by two Mo planes.
It has an orthorhombic unit cell, with each Mo atom
sitting in an octahedral environment of O atoms.
It is an indirect gap semiconductor, even in the 2D form, with a gap between
the conduction band at $\Gamma$ and the
valence band at $M$.\cite{molina2015centimeter}

\begin{figure}[t!]
 \centering
                \includegraphics[width=.5\textwidth]{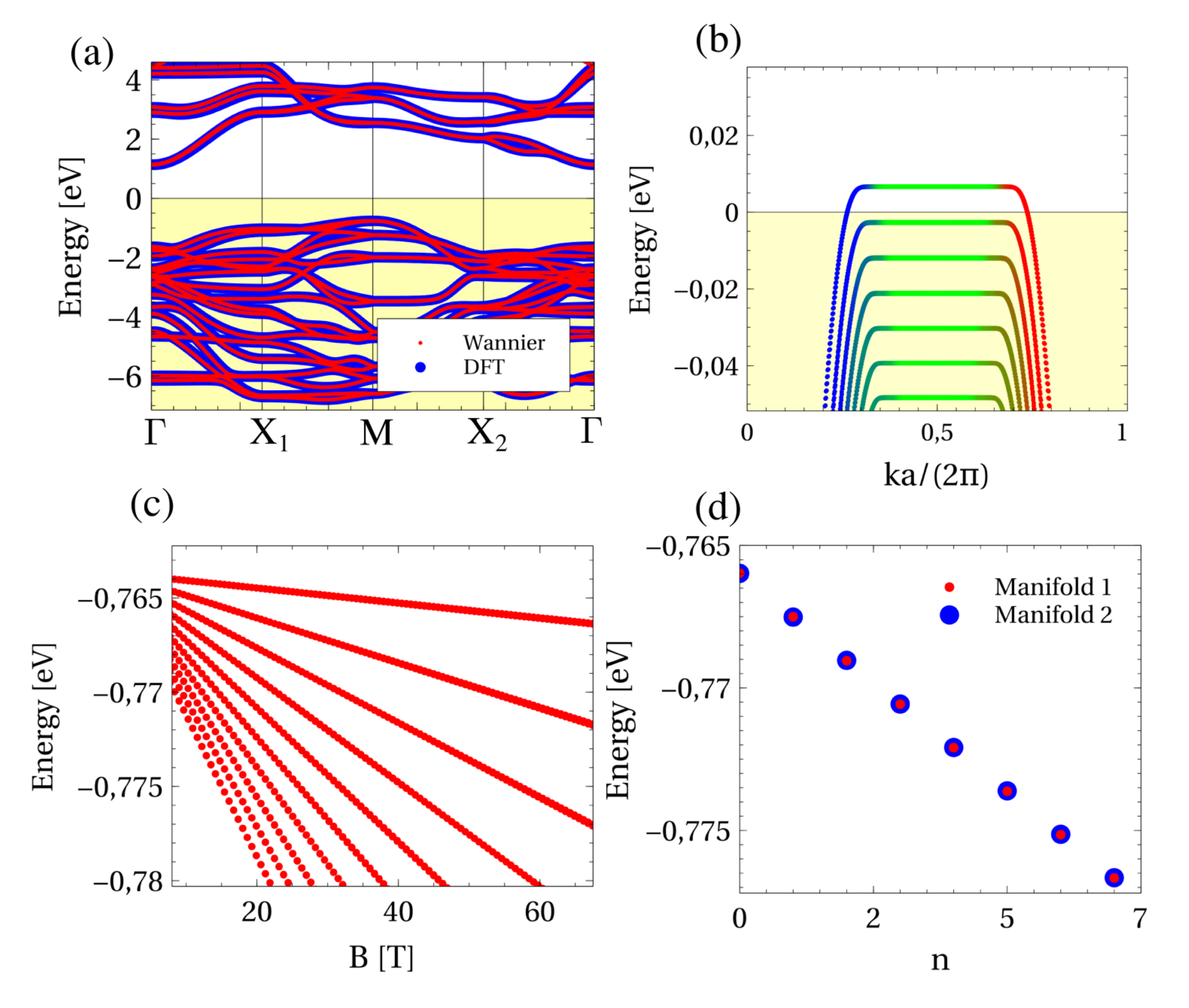}

\caption{ (a) Comparison of the band structure of monolayer
MoO$_3$ as obtained by DFT and the Wannier
tight binding Hamiltonian, showing
a maximum in the valence band at $M$,
with two spinless degenerate bands. (b) Band structure of
MoO$_3$
nanoribbon of 40 nm thickness
in the quantum Hall regime ($B = 115$ T), focusing on the valence band,
showing a conventional Landau level spectra. (c) Scaling of the
Landau levels energy with the magnetic field, and with
the Landau index (d) at a fixed magnetic field, showing that
the two sets of Landau levels are degenerate in energy. Ribbon
thickness in (c,d) is 160 nm
}
\label{fig:MoO3}
\end{figure}

The low energy properties of this compound can be captured
taking into account the d orbitals of Mo and the p orbitals
of O, giving rise to $28\times28$ tight binding Hamiltonian.
In the following,
we will focus on the  Landau Levels of  the valence band,
which corresponds to the parabolic band located at $M$. At that point,
two bands coexist, which correspond to the two different Mo layers in the
unit cell, each one
giving rise to one set of Landau levels. 
The  MoO$_3$ Landau level spectrum, shown in Fig. \ref{fig:MoO3}b, 
is the one expected for Schrodinger quasiparticles, except for the twofold orbital degeneracy associated
to the layer index. 
The degeneracy remains as we ramp the magnetic field
 (Fig. \ref{fig:MoO3}c), and for the different
Landau indexes (Fig. \ref{fig:MoO3}d).  The degeneracy could be lifted upon application
of a  perpendicular electric field or, more interestingly, if electronic
order arises due to electron-electron
interaction.

\section{Landau levels as a measure of Dirac-ness}

We now briefly discuss the concept of Dirac-ness and we use a recently 
proposed method to quantify it.\cite{goerbig2014measure}  
There is no doubt that quasiparticles in graphene behave as Dirac electrons. 
However, when  a given 2D crystal has a gap, things are less clear as the energy dispersion of two Schrodinger bands
is very similar to the spectrum of a gapped Dirac equation. 
A simple way
to measure the Dirac-ness of a band structure consists on
looking at the evolution of the zero Landau level with
magnetic field.\cite{goerbig2014measure} This idea is based on the
fact that for Schrodinger fermions, the energy of the Landau levels
follow $E_n = \Delta + (\hbar e B/m)(n+1/2)$ whereas for Dirac massive fermions
in the large mass regime $E_n \approx \Delta + (\hbar e B/m)n$, where $\Delta$
in the energy off-set of the band and m the effective mass.
The previous expression can be generalized into 
\begin{equation}
E_n = \Delta + (\hbar e B/m)(n+\gamma)
\label{diracness}
\end{equation}
with $\gamma=0$ for Dirac and $\gamma=1/2$ for Schrodinger. 
Therefore, a key feature of massive Dirac fermions is that
the energy of the zero Landau level is independent of the magnetic field,
showing a flat evolution of the Landau level energy versus
magnetic field. On the contrary, conventional Schrodinger fermions
will have all their Landau levels dependent on the magnetic field,
so that no flat evolution is observed. 

The Dirac-ness can be easily observed
by checking the Landau levels versus magnetic field
as obtained for the
different materials. In particular, graphene (Fig. \ref{fig:graphene}c),
boron nitride (Fig. \ref{fig:BN}c) and MoS$_2$ (Fig. \ref{fig:MoS2}c) show
a zero Landau level whose energy is independent
of the magnetic field, and thereby they are
 Dirac materials. In  contrast,  black phosphorus
(Fig. \ref{fig:P}c), InSe (Fig. \ref{fig:InSe}c) and
MoO$_3$ (Fig. \ref{fig:MoO3}c) show a first Landau level with non-zero
slope. 
\redmark{
By fitting the first four Landau levels
at the different magnetic fields shown in Figs.
\ref{fig:graphene}
\ref{fig:BN}
\ref{fig:MoS2}
\ref{fig:P}
\ref{fig:InSe}
\ref{fig:MoO3}: 
to Eq. \ref{diracness}, we
obtain the values shown in Table. \ref{table}.
}

\begin{center}
\begin{table}
\begin{tabular}{|c|c|c|}
\hline
Material & $\gamma$ \\
\hline
graphene & 0.0 \\
BN, valley $K'$ (valence) & 0.024 \\
BN, valley $K$ (valence) & -0.03 \\
MoS$_{2}$,valley $K'$ (conduction)& 0.017 \\
MoS$_{2}$, valley $K$ (conduction) & -0.03 \\
Black Phosphorus (conduction) & 0.499 \\
InSe (conduction) & 0.496 \\
MoO$_3$ (valence) & 0.5 \\
\hline
\end{tabular}
\caption{Values of $\gamma$ for the different materials obtained
by fitting the LL versus magnetic field to Eq. \ref{diracness}}
\label{table}
\end{table}
\end{center}


\section{Conclusions}

We have shown that
using the Wannierization
procedure
a faithful
tight-binding representation
of the DFT Hamiltonian
can be obtained for several
2D materials.
We have shown how the addition of a Peierls phase into the Wannier Hamiltonian
permits to compute the Landau Level  spectrum, both
for bulk and edge states,
of a variety of  2D materials, including graphene, BN, MoS$_2$,
Black Phosphorous, Indium Selenide and MoO$_3$.
 The method is particularly suitable for  systems lacking a reliable tight-binding model, or for which the derivation
 of an effective mass $kp$ Hamiltonian is complicated or not available.   
We have also shown that by analyzing  the evolution of the Landau level
spectra, the Dirac-ness of the band structure can be determined,
yielding a simple tool to identify materials with Dirac physics.

\section*{Acknowledgments}

 JFR acknowledges  financial supported by MEC-Spain (FIS2013-47328-C2-2-P) 
  and Generalitat Valenciana (ACOMP/2010/070), Prometeo. This work has
been financially supported in part by FEDER funds. We thank
B. Amorim and N. Garcia-Martinez for fruitful discussions.  We acknowledge
financial support by Marie-Curie-ITN 607904-SPINOGRAPH. J. L. Lado thanks
the hospitality of the Departamento de Fisica Aplicada
at the Universidad de Alicante. 
This work was supported by National Funds through the Portuguese Foundation for
Science and Technology (FCT) in the framework of the Strategic Funding
UID/FIS/04650/2013¿ and through project PTDC/FIS-NAN/3668/2014

\appendix

\section{Computational details}

 The starting point is density functional calculations,
performed with Quantum
Espresso, for the unit cell of the desired 2D crystal. We
use PBE\cite{PhysRevLett.77.3865} 
functional and the PAW pseudopotentials\cite{lejaeghere2016reproducibility}
for structural relaxation.
Wannierization
\cite{PhysRevB.56.12847,mostofi2008wannier90,PhysRevB.65.035109,PhysRevB.69.035108,RevModPhys.84.1419}
 is performed in $20\times20\times1$ kmesh,
over the non relativistic calculation
with PAW pseudopotentials,
except for the case of relativistic MoS$_2$ where we used
norm conserving pseudopotentials.
With the Wannier Hamiltonian for bulk, the tight
binding Hamiltonian for the ribbon is created
by taking the relevant tight binding parameters of the bulk Hamiltonian
for each cell replica, considering hoppings
up to third neighbouring cells. 

The scaling of the Landau levels with magnetic field is calculated
first by determining the position in reciprocal space of the
flat Landau bands (in the case of graphene, BN and MoS$_2$
two different regions).
The diagonalization of the quantum Hall slab
is performed in that kpoint, retaining only
those eigenvalues whose eigenfunctions are located in the bulk
to throw away edge states and dangling bond eigenvalues
(see Figs. \ref{fig:MoS2}b, \ref{fig:InSe}b).
This procedure allows to study very wide ribbons, since
the tight binding Hamiltonian is highly sparse and a few eigenvalues
around the Fermi energy can be efficiently
calculated with ARPACK.\cite{lehoucq1998arpack}

\bibliographystyle{ieeetr}
\bibliography{biblio}{}

\begin{thebibliography}{10}

\bibitem{landau1962}
L.~Landau and E.~Lifshitz, ``Quantum mechanics (non-relativistic theory) course
  on theoretical physics, vol. 3, pergamon,'' 1962.

\bibitem{AndoRMP}
T.~Ando, A.~B. Fowler, and F.~Stern, ``Electronic properties of two-dimensional
  systems,'' {\em Rev. Mod. Phys.}, vol.~54, pp.~437--672, Apr 1982.

\bibitem{novoselov2005}
K.~Novoselov, A.~K. Geim, S.~Morozov, D.~Jiang, M.~Katsnelson, I.~Grigorieva,
  S.~Dubonos, and A.~Firsov, ``Two-dimensional gas of massless dirac fermions
  in graphene,'' {\em nature}, vol.~438, no.~7065, pp.~197--200, 2005.

\bibitem{Zhang2005}
Y.~Zhang, Y.-W. Tan, H.~L. Stormer, and P.~Kim, ``Experimental observation of
  the quantum hall effect and berry's phase in graphene,'' {\em Nature},
  vol.~438, no.~7065, pp.~201--204, 2005.

\bibitem{gillgren2014gate}
N.~Gillgren, D.~Wickramaratne, Y.~Shi, T.~Espiritu, J.~Yang, J.~Hu, J.~Wei,
  X.~Liu, Z.~Mao, K.~Watanabe, {\em et~al.}, ``Gate tunable quantum
  oscillations in air-stable and high mobility few-layer phosphorene
  heterostructures,'' {\em 2D Materials}, vol.~2, no.~1, p.~011001, 2014.

\bibitem{li2015quantum}
L.~Li, G.~J. Ye, V.~Tran, R.~Fei, G.~Chen, H.~Wang, J.~Wang, K.~Watanabe,
  T.~Taniguchi, L.~Yang, {\em et~al.}, ``Quantum oscillations in a
  two-dimensional electron gas in black phosphorus thin films,'' {\em Nature
  nanotechnology}, vol.~10, no.~7, pp.~608--613, 2015.

\bibitem{tayari2015two}
V.~Tayari, N.~Hemsworth, I.~Fakih, A.~Favron, E.~Gaufr{\`e}s, G.~Gervais,
  R.~Martel, and T.~Szkopek, ``Two-dimensional magnetotransport in a black
  phosphorus naked quantum well,'' {\em Nature communications}, vol.~6, 2015.

\bibitem{li2016}
L.~Li, F.~Yang, G.~J. Ye, Z.~Zhang, Z.~Zhu, W.~Lou, X.~Zhou, L.~Li,
  K.~Watanabe, T.~Taniguchi, {\em et~al.}, ``Quantum hall effect in black
  phosphorus two-dimensional electron system,'' {\em Nature nanotechnology},
  2016.

\bibitem{wu2015}
Z.~Wu, S.~Xu, H.~Lu, G.-B. Liu, A.~Khamoshi, T.~Han, Y.~Wu, J.~Lin, G.~Long,
  Y.~He, {\em et~al.}, ``Observation of valley zeeman and quantum hall effects
  at q valley of few-layer transition metal disulfides,'' {\em arXiv preprint
  arXiv:1511.00077}, 2015.

\bibitem{PhysRevLett.116.086601}
B.~Fallahazad, H.~C.~P. Movva, K.~Kim, S.~Larentis, T.~Taniguchi, K.~Watanabe,
  S.~K. Banerjee, and E.~Tutuc, ``Shubnikov\char21{}de haas oscillations of
  high-mobility holes in monolayer and bilayer ${\mathrm{wse}}_{2}$: Landau
  level degeneracy, effective mass, and negative compressibility,'' {\em Phys.
  Rev. Lett.}, vol.~116, p.~086601, Feb 2016.

\bibitem{zheng2002}
Y.~Zheng and T.~Ando, ``Hall conductivity of a two-dimensional graphite
  system,'' {\em Physical Review B}, vol.~65, no.~24, p.~245420, 2002.

\bibitem{Niu2013}
X.~Li, F.~Zhang, and Q.~Niu, ``Unconventional quantum hall effect and tunable
  spin hall effect in dirac materials: Application to an isolated
  ${\mathrm{mos}}_{2}$ trilayer,'' {\em Phys. Rev. Lett.}, vol.~110, p.~066803,
  Feb 2013.

\bibitem{pereira2015}
J.~M. Pereira and M.~I. Katsnelson, ``Landau levels of single-layer and bilayer
  phosphorene,'' {\em Phys. Rev. B}, vol.~92, p.~075437, Aug 2015.

\bibitem{yuan2015}
S.~Yuan, M.~I. Katsnelson, and R.~Rold{\'a}n, ``Quantum hall effect in biased
  black phosphorus,'' {\em arXiv preprint arXiv:1512.06345}, 2015.

\bibitem{saito1998physical}
R.~Saito, G.~Dresselhaus, M.~S. Dresselhaus, {\em et~al.}, {\em Physical
  properties of carbon nanotubes}, vol.~35.
\newblock World Scientific, 1998.

\bibitem{katsnelson2012}
M.~Katsnelson, {\em Graphene: carbon in two dimensions}.
\newblock Cambridge University Press, 2012.

\bibitem{Halperin1982}
B.~I. Halperin, ``Quantized hall conductance, current-carrying edge states, and
  the existence of extended states in a two-dimensional disordered potential,''
  {\em Phys. Rev. B}, vol.~25, pp.~2185--2190, Feb 1982.

\bibitem{breyfertig}
L.~Brey and H.~A. Fertig, ``Edge states and the quantized hall effect in
  graphene,'' {\em Phys. Rev. B}, vol.~73, p.~195408, May 2006.

\bibitem{RevModPhys.81.109}
A.~H. Castro~Neto, F.~Guinea, N.~M.~R. Peres, K.~S. Novoselov, and A.~K. Geim,
  ``The electronic properties of graphene,'' {\em Rev. Mod. Phys.}, vol.~81,
  pp.~109--162, Jan 2009.

\bibitem{Lado2013}
J.~L. Lado, J.~W. Gonz{\'a}lez, and J.~Fern{\'a}ndez-Rossier, ``Quantum hall
  effect in gapped graphene heterojunctions,'' {\em Physical Review B},
  vol.~88, no.~3, p.~035448, 2013.

\bibitem{PhysRevB.56.12847}
N.~Marzari and D.~Vanderbilt, ``Maximally localized generalized wannier
  functions for composite energy bands,'' {\em Phys. Rev. B}, vol.~56,
  pp.~12847--12865, Nov 1997.

\bibitem{mostofi2008wannier90}
A.~A. Mostofi, J.~R. Yates, Y.-S. Lee, I.~Souza, D.~Vanderbilt, and N.~Marzari,
  ``wannier90: A tool for obtaining maximally-localised wannier functions,''
  {\em Computer physics communications}, vol.~178, no.~9, pp.~685--699, 2008.

\bibitem{PhysRevB.65.035109}
I.~Souza, N.~Marzari, and D.~Vanderbilt, ``Maximally localized wannier
  functions for entangled energy bands,'' {\em Phys. Rev. B}, vol.~65,
  p.~035109, Dec 2001.

\bibitem{PhysRevB.69.035108}
A.~Calzolari, N.~Marzari, I.~Souza, and M.~Buongiorno~Nardelli, ``\textit{Ab
  initio} transport properties of nanostructures from maximally localized
  wannier functions,'' {\em Phys. Rev. B}, vol.~69, p.~035108, Jan 2004.

\bibitem{RevModPhys.84.1419}
N.~Marzari, A.~A. Mostofi, J.~R. Yates, I.~Souza, and D.~Vanderbilt,
  ``Maximally localized wannier functions: Theory and applications,'' {\em Rev.
  Mod. Phys.}, vol.~84, pp.~1419--1475, Oct 2012.

\bibitem{PhysRevB.74.235111}
T.~Thonhauser and D.~Vanderbilt, ``Insulator/chern-insulator transition in the
  haldane model,'' {\em Phys. Rev. B}, vol.~74, p.~235111, Dec 2006.

\bibitem{PhysRevB.88.245436}
K.~Ko\ifmmode~\acute{s}\else \'{s}\fi{}mider, J.~W. Gonz\'alez, and
  J.~Fern\'andez-Rossier, ``Large spin splitting in the conduction band of
  transition metal dichalcogenide monolayers,'' {\em Phys. Rev. B}, vol.~88,
  p.~245436, Dec 2013.

\bibitem{PhysRevB.87.235420}
L.~Chen, Z.~F. Wang, and F.~Liu, ``Robustness of two-dimensional topological
  insulator states in bilayer bismuth against strain and electrical field,''
  {\em Phys. Rev. B}, vol.~87, p.~235420, Jun 2013.

\bibitem{lado2016dirac}
J.~Lado and V.~Pardo, ``Dirac topological insulator in the dz2 manifold of a
  honeycomb oxide,'' {\em arXiv preprint arXiv:1604.05554}, 2016.

\bibitem{jung2013}
J.~Jung and A.~H. MacDonald, ``Tight-binding model for graphene
  $\ensuremath{\pi}$-bands from maximally localized wannier functions,'' {\em
  Phys. Rev. B}, vol.~87, p.~195450, May 2013.

\bibitem{PhysRevB.92.161115}
H.~Huang, Z.~Liu, H.~Zhang, W.~Duan, and D.~Vanderbilt, ``Emergence of a
  chern-insulating state from a semi-dirac dispersion,'' {\em Phys. Rev. B},
  vol.~92, p.~161115, Oct 2015.

\bibitem{assmann2015woptic}
E.~Assmann, P.~Wissgott, J.~Kune{\v{s}}, A.~Toschi, P.~Blaha, and K.~Held,
  ``woptic: optical conductivity with wannier functions and adaptive k-mesh
  refinement,'' {\em Computer Physics Communications}, 2015.

\bibitem{pizzi2014boltzwann}
G.~Pizzi, D.~Volja, B.~Kozinsky, M.~Fornari, and N.~Marzari, ``Boltzwann: A
  code for the evaluation of thermoelectric and electronic transport properties
  with a maximally-localized wannier functions basis,'' {\em Computer Physics
  Communications}, vol.~185, no.~1, pp.~422--429, 2014.

\bibitem{PhysRevB.74.125120}
F.~Lechermann, A.~Georges, A.~Poteryaev, S.~Biermann, M.~Posternak,
  A.~Yamasaki, and O.~K. Andersen, ``Dynamical mean-field theory using wannier
  functions: A flexible route to electronic structure calculations of strongly
  correlated materials,'' {\em Phys. Rev. B}, vol.~74, p.~125120, Sep 2006.

\bibitem{PhysRevB.88.165427}
A.~Kretinin, G.~L. Yu, R.~Jalil, Y.~Cao, F.~Withers, A.~Mishchenko, M.~I.
  Katsnelson, K.~S. Novoselov, A.~K. Geim, and F.~Guinea, ``Quantum capacitance
  measurements of electron-hole asymmetry and next-nearest-neighbor hopping in
  graphene,'' {\em Phys. Rev. B}, vol.~88, p.~165427, Oct 2013.

\bibitem{Note1}
As opposed to being localized in the edge.

\bibitem{gorbachev2011hunting}
R.~V. Gorbachev, I.~Riaz, R.~R. Nair, R.~Jalil, L.~Britnell, B.~D. Belle, E.~W.
  Hill, K.~S. Novoselov, K.~Watanabe, T.~Taniguchi, {\em et~al.}, ``Hunting for
  monolayer boron nitride: optical and raman signatures,'' {\em Small}, vol.~7,
  no.~4, pp.~465--468, 2011.

\bibitem{alem2009atomically}
N.~Alem, R.~Erni, C.~Kisielowski, M.~D. Rossell, W.~Gannett, and A.~Zettl,
  ``Atomically thin hexagonal boron nitride probed by ultrahigh-resolution
  transmission electron microscopy,'' {\em Physical Review B}, vol.~80, no.~15,
  p.~155425, 2009.

\bibitem{young2014tunable}
A.~F. Young, J.~Sanchez-Yamagishi, B.~Hunt, S.~H. Choi, K.~Watanabe,
  T.~Taniguchi, R.~Ashoori, and P.~Jarillo-Herrero, ``Tunable symmetry breaking
  and helical edge transport in a graphene quantum spin hall state,'' {\em
  Nature}, vol.~505, no.~7484, pp.~528--532, 2014.

\bibitem{PhysRevB.93.115441}
M.~Gurram, S.~Omar, S.~Zihlmann, P.~Makk, C.~Sch\"onenberger, and B.~J. van
  Wees, ``Spin transport in fully hexagonal boron nitride encapsulated
  graphene,'' {\em Phys. Rev. B}, vol.~93, p.~115441, Mar 2016.

\bibitem{doi:10.1021/acsnano.5b01341}
G.-H. Lee, X.~Cui, Y.~D. Kim, G.~Arefe, X.~Zhang, C.-H. Lee, F.~Ye,
  K.~Watanabe, T.~Taniguchi, P.~Kim, and J.~Hone, ``Highly stable, dual-gated
  mos2 transistors encapsulated by hexagonal boron nitride with
  gate-controllable contact, resistance, and threshold voltage,'' {\em ACS
  Nano}, vol.~9, no.~7, pp.~7019--7026, 2015.
\newblock PMID: 26083310.

\bibitem{britnell2012atomically}
L.~Britnell, R.~V. Gorbachev, R.~Jalil, B.~D. Belle, F.~Schedin, M.~I.
  Katsnelson, L.~Eaves, S.~V. Morozov, A.~S. Mayorov, N.~M. Peres, {\em
  et~al.}, ``Atomically thin boron nitride: a tunnelling barrier for graphene
  devices,'' {\em arXiv preprint arXiv:1202.0735}, 2012.

\bibitem{cassabois2015hexagonal}
G.~Cassabois, P.~Valvin, and B.~Gil, ``Hexagonal boron nitride is an indirect
  bandgap semiconductor,'' {\em arXiv preprint arXiv:1512.02962}, 2015.

\bibitem{koshino2010}
M.~Koshino and T.~Ando, ``Anomalous orbital magnetism in dirac-electron
  systems: Role of pseudospin paramagnetism,'' {\em Phys. Rev. B}, vol.~81,
  p.~195431, May 2010.

\bibitem{wang2012}
Q.~H. Wang, K.~Kalantar-Zadeh, A.~Kis, J.~N. Coleman, and M.~S. Strano,
  ``Electronics and optoelectronics of two-dimensional transition metal
  dichalcogenides,'' {\em Nature nanotechnology}, vol.~7, no.~11, pp.~699--712,
  2012.

\bibitem{zahid2013generic}
F.~Zahid, L.~Liu, Y.~Zhu, J.~Wang, and H.~Guo, ``A generic tight-binding model
  for monolayer, bilayer and bulk mos2,'' {\em AIP Advances}, vol.~3, no.~5,
  p.~052111, 2013.

\bibitem{cappelluti2013tight}
E.~Cappelluti, R.~Rold{\'a}n, J.~Silva-Guill{\'e}n, P.~Ordej{\'o}n, and
  F.~Guinea, ``Tight-binding model and direct-gap/indirect-gap transition in
  single-layer and multilayer mos 2,'' {\em Physical Review B}, vol.~88, no.~7,
  p.~075409, 2013.

\bibitem{ridolfi2015tight}
E.~Ridolfi, D.~Le, T.~Rahman, E.~Mucciolo, and C.~Lewenkopf, ``A tight-binding
  model for mos2 monolayers,'' {\em Journal of Physics: Condensed Matter},
  vol.~27, no.~36, p.~365501, 2015.

\bibitem{PhysRevB.88.085433}
G.-B. Liu, W.-Y. Shan, Y.~Yao, W.~Yao, and D.~Xiao, ``Three-band tight-binding
  model for monolayers of group-vib transition metal dichalcogenides,'' {\em
  Phys. Rev. B}, vol.~88, p.~085433, Aug 2013.

\bibitem{kormanyos2015k}
A.~Korm{\'a}nyos, G.~Burkard, M.~Gmitra, J.~Fabian, V.~Z{\'o}lyomi, N.~D.
  Drummond, and V.~Fal’ko, ``k{\textperiodcentered} p theory for
  two-dimensional transition metal dichalcogenide semiconductors,'' {\em 2D
  Materials}, vol.~2, no.~2, p.~022001, 2015.

\bibitem{liu2015electronic}
G.-B. Liu, D.~Xiao, Y.~Yao, X.~Xu, and W.~Yao, ``Electronic structures and
  theoretical modelling of two-dimensional group-vib transition metal
  dichalcogenides,'' {\em Chemical Society Reviews}, vol.~44, no.~9,
  pp.~2643--2663, 2015.

\bibitem{PhysRevB.93.035406}
M.~Tahir, P.~Vasilopoulos, and F.~M. Peeters, ``Quantum magnetotransport
  properties of a ${\text{mos}}_{2}$ monolayer,'' {\em Phys. Rev. B}, vol.~93,
  p.~035406, Jan 2016.

\bibitem{PhysRevB.88.125438}
F.~Rose, M.~O. Goerbig, and F.~Pi\'echon, ``Spin- and valley-dependent
  magneto-optical properties of mos${}_{2}$,'' {\em Phys. Rev. B}, vol.~88,
  p.~125438, Sep 2013.

\bibitem{xiao2012}
D.~Xiao, G.-B. Liu, W.~Feng, X.~Xu, and W.~Yao, ``Coupled spin and valley
  physics in monolayers of mos 2 and other group-vi dichalcogenides,'' {\em
  Physical Review Letters}, vol.~108, no.~19, p.~196802, 2012.

\bibitem{PhysRevB.90.045427}
R.-L. Chu, X.~Li, S.~Wu, Q.~Niu, W.~Yao, X.~Xu, and C.~Zhang,
  ``Valley-splitting and valley-dependent inter-landau-level optical
  transitions in monolayer ${\mathrm{mos}}_{2}$ quantum hall systems,'' {\em
  Phys. Rev. B}, vol.~90, p.~045427, Jul 2014.

\bibitem{PhysRevLett.114.037401}
D.~MacNeill, C.~Heikes, K.~F. Mak, Z.~Anderson, A.~Korm\'anyos, V.~Z\'olyomi,
  J.~Park, and D.~C. Ralph, ``Breaking of valley degeneracy by magnetic field
  in monolayer ${\mathrm{mose}}_{2}$,'' {\em Phys. Rev. Lett.}, vol.~114,
  p.~037401, Jan 2015.

\bibitem{PhysRevB.91.075433}
H.~Rostami and R.~Asgari, ``Valley zeeman effect and spin-valley polarized
  conductance in monolayer ${\text{mos}}_{2}$ in a perpendicular magnetic
  field,'' {\em Phys. Rev. B}, vol.~91, p.~075433, Feb 2015.

\bibitem{PhysRevB.87.235109}
R.~Sakuma, ``Symmetry-adapted wannier functions in the maximal localization
  procedure,'' {\em Phys. Rev. B}, vol.~87, p.~235109, Jun 2013.

\bibitem{PhysRevB.80.235431}
M.~Gmitra, S.~Konschuh, C.~Ertler, C.~Ambrosch-Draxl, and J.~Fabian,
  ``Band-structure topologies of graphene: Spin-orbit coupling effects from
  first principles,'' {\em Phys. Rev. B}, vol.~80, p.~235431, Dec 2009.

\bibitem{PhysRevB.82.245412}
S.~Konschuh, M.~Gmitra, and J.~Fabian, ``Tight-binding theory of the spin-orbit
  coupling in graphene,'' {\em Phys. Rev. B}, vol.~82, p.~245412, Dec 2010.

\bibitem{PhysRevB.74.165310}
H.~Min, J.~E. Hill, N.~A. Sinitsyn, B.~R. Sahu, L.~Kleinman, and A.~H.
  MacDonald, ``Intrinsic and rashba spin-orbit interactions in graphene
  sheets,'' {\em Phys. Rev. B}, vol.~74, p.~165310, Oct 2006.

\bibitem{PhysRevB.75.041401}
Y.~Yao, F.~Ye, X.-L. Qi, S.-C. Zhang, and Z.~Fang, ``Spin-orbit gap of
  graphene: First-principles calculations,'' {\em Phys. Rev. B}, vol.~75,
  p.~041401, Jan 2007.

\bibitem{PhysRevLett.95.137205}
T.~Thonhauser, D.~Ceresoli, D.~Vanderbilt, and R.~Resta, ``Orbital
  magnetization in periodic insulators,'' {\em Phys. Rev. Lett.}, vol.~95,
  p.~137205, Sep 2005.

\bibitem{resta2010electrical}
R.~Resta, ``Electrical polarization and orbital magnetization: the modern
  theories,'' {\em Journal of Physics: Condensed Matter}, vol.~22, no.~12,
  p.~123201, 2010.

\bibitem{PhysRevB.74.024408}
D.~Ceresoli, T.~Thonhauser, D.~Vanderbilt, and R.~Resta, ``Orbital
  magnetization in crystalline solids: Multi-band insulators, chern insulators,
  and metals,'' {\em Phys. Rev. B}, vol.~74, p.~024408, Jul 2006.

\bibitem{castellanos2014isolation}
A.~Castellanos-Gomez, L.~Vicarelli, E.~Prada, J.~O. Island,
  K.~Narasimha-Acharya, S.~I. Blanter, D.~J. Groenendijk, M.~Buscema, G.~A.
  Steele, J.~Alvarez, {\em et~al.}, ``Isolation and characterization of
  few-layer black phosphorus,'' {\em 2D Materials}, vol.~1, no.~2, p.~025001,
  2014.

\bibitem{takao1981electronic}
Y.~Takao and A.~Morita, ``Electronic structure of black phosphorus: tight
  binding approach,'' {\em Physica B+ C}, vol.~105, no.~1-3, pp.~93--98, 1981.

\bibitem{rudenko2014quasiparticle}
A.~N. Rudenko and M.~I. Katsnelson, ``Quasiparticle band structure and
  tight-binding model for single-and bilayer black phosphorus,'' {\em Physical
  Review B}, vol.~89, no.~20, p.~201408, 2014.

\bibitem{rudenko2015toward}
A.~Rudenko, S.~Yuan, and M.~Katsnelson, ``Toward a realistic description of
  multilayer black phosphorus: From g w approximation to large-scale
  tight-binding simulations,'' {\em Physical Review B}, vol.~92, no.~8,
  p.~085419, 2015.

\bibitem{zhou2015landau}
X.~Zhou, R.~Zhang, J.~Sun, Y.~Zou, D.~Zhang, W.~Lou, F.~Cheng, G.~Zhou,
  F.~Zhai, and K.~Chang, ``Landau levels and magneto-transport property of
  monolayer phosphorene,'' {\em Scientific reports}, vol.~5, 2015.

\bibitem{debbichi2015two}
L.~Debbichi, O.~Eriksson, and S.~Leb{\`e}gue, ``Two-dimensional indium
  selenides compounds: An ab initio study,'' {\em The journal of physical
  chemistry letters}, vol.~6, no.~15, pp.~3098--3103, 2015.

\bibitem{lauth2016solution}
J.~Lauth, F.~E. Gorris, M.~Samadi~Khoshkhoo, T.~Chassé, W.~Friedrich,
  V.~Lebedeva, A.~Meyer, C.~Klinke, A.~Kornowski, M.~Scheele, {\em et~al.},
  ``Solution-processed two-dimensional ultrathin inse nanosheets,'' {\em
  Chemistry of Materials}, vol.~28, no.~6, pp.~1728--1736, 2016.

\bibitem{lei2014evolution}
S.~Lei, L.~Ge, S.~Najmaei, A.~George, R.~Kappera, J.~Lou, M.~Chhowalla,
  H.~Yamaguchi, G.~Gupta, R.~Vajtai, {\em et~al.}, ``Evolution of the
  electronic band structure and efficient photo-detection in atomic layers of
  inse,'' {\em ACS nano}, vol.~8, no.~2, pp.~1263--1272, 2014.

\bibitem{electroInSe}
N.~Balakrishnan, Z.~R. Kudrynskyi, M.~W. Fay, G.~W. Mudd, S.~A. Svatek,
  O.~Makarovsky, Z.~D. Kovalyuk, L.~Eaves, P.~H. Beton, and A.~Patanè, ``Room
  temperature electroluminescence from mechanically formed van der waals
  iii–vi homojunctions and heterojunctions,'' {\em Advanced Optical
  Materials}, vol.~2, no.~11, pp.~1064--1069, 2014.

\bibitem{brotons2016nanotexturing}
M.~Brotons-Gisbert, D.~Andres-Penares, J.~Suh, F.~Hidalgo, R.~Abargues, P.~J.
  Rodr{\'\i}guez-Cant{\'o}, A.~Segura, A.~Cros, G.~Tobias, E.~Canadell, {\em
  et~al.}, ``Nanotexturing to enhance photoluminescent response of atomically
  thin indium selenide with highly tunable band gap,'' {\em Nano Letters},
  2016.

\bibitem{kalantar2010synthesis}
K.~Kalantar-Zadeh, J.~Tang, M.~Wang, K.~L. Wang, A.~Shailos, K.~Galatsis,
  R.~Kojima, V.~Strong, A.~Lech, W.~Wlodarski, {\em et~al.}, ``Synthesis of
  nanometre-thick moo 3 sheets,'' {\em Nanoscale}, vol.~2, no.~3, pp.~429--433,
  2010.

\bibitem{molina2015centimeter}
A.~J. Molina-Mendoza, J.~L. Lado, J.~Island, M.~A. Ni{\~n}o, L.~Aballe,
  M.~Foerster, F.~Y. Bruno, H.~S. van~der Zant, G.~Rubio-Bollinger,
  N.~Agra{\"\i}t, {\em et~al.}, ``Centimeter-scale synthesis of ultrathin
  layered moo3 by van der waals epitaxy,'' {\em arXiv preprint
  arXiv:1512.04355}, 2015.

\bibitem{goerbig2014measure}
M.~Goerbig, G.~Montambaux, {\em et~al.}, ``Measure of diracness in
  two-dimensional semiconductors,'' {\em EPL (Europhysics Letters)}, vol.~105,
  no.~5, p.~57005, 2014.

\bibitem{PhysRevLett.77.3865}
J.~P. Perdew, K.~Burke, and M.~Ernzerhof, ``Generalized gradient approximation
  made simple,'' {\em Phys. Rev. Lett.}, vol.~77, pp.~3865--3868, Oct 1996.

\bibitem{lejaeghere2016reproducibility}
K.~Lejaeghere, G.~Bihlmayer, T.~Bj{\"o}rkman, P.~Blaha, S.~Bl{\"u}gel, V.~Blum,
  D.~Caliste, I.~E. Castelli, S.~J. Clark, A.~Dal~Corso, {\em et~al.},
  ``Reproducibility in density functional theory calculations of solids,'' {\em
  Science}, vol.~351, no.~6280, p.~aad3000, 2016.

\bibitem{lehoucq1998arpack}
R.~B. Lehoucq, D.~C. Sorensen, and C.~Yang, {\em ARPACK users' guide: solution
  of large-scale eigenvalue problems with implicitly restarted Arnoldi
  methods}, vol.~6.
\newblock Siam, 1998.

\end{thebibliography}

\end{document}